\documentclass[dvips]{acta}
\usepackage{supertabular,lscape,epsfig,hyperref}
\usepackage{amssymb}
\usepackage{amsmath}

\SetPages{1}{18}

\SetVol{71}{2021}

\begin{document}

\begin{Titlepage}
\Title{Photometric and spectroscopic investigations of the Galactic field RRc candidates V764\,Mon and HY\,Com}
\Author{J. M. Benk\H{o}$^{1,2}$, \'A. S\'odor$^{1,2}$ and A. P\'al$^1$}
{$^{1}$Konkoly Observatory, Research Centre for Astronomy and Earth Sciences, ELKH,\\
Konkoly Thege Mikl\'os \'ut 15-17, H-1121 Budapest, Hungary\\
$^{2}$MTA CSFK Lend\"ulet Near-Field Cosmology Research Group\\
e-mail:benko@konkoly.hu}

\Received{January 29, 2021}
\end{Titlepage}

\Abstract{By analysing photometric and spectroscopic time series in this paper, we show that the pulsator V764\,Mon, assumed 
to be the brightest RR Lyrae star in the sky, is in fact a rapidly rotating $\delta$\,Scuti star with an unusually long 
dominant period ($P_1=0.29$~d). 
Our spectroscopy confirmed the discovery of the Gaia satellite about the binarity of V764\,Mon.
In the case of HY\,Com,
a `bona fide' RRc star, we present its first complete radial velocity curve. Additionally, we found that the 
star continues its strong phase variation reported before.
}
{stars: oscillations -- stars: variables: RR\,Lyrae
-- stars: variables: $\delta$\,Scuti -- methods: data analysis -- space vehicles}

\section{Introduction}

The catalogue of Galactic RR Lyrae stars (Maintz 2005) 
contains only eight first-overtone (RRc) pulsators in the sky
that are visually brighter than 10 mag at maximum light.
The brightest one is V764\,Mon (= HD 61176, A5IV, $V_{\mathrm{max}}=7.13$~mag).
The variable nature of the star was discovered by the {\it Hipparcos} satellite (ESA 1997). 
On the basis of its period ($P=0.290$~d), it was classified as an RRc candidate, as such, not just the 
brightest RRc star, but it would even be the brightest of all known RR Lyrae, because
it is also brighter than the brightest fundamental-mode RR Lyrae pulsator, the namesake of the class, RR\,Lyr
 ($V_{\mathrm{max}}=7.45$~mag) itself.
It is therefore surprising how neglected this star is. Up to now no dedicated study targeted it.

Kervella et al. (2019a,b) enumerate V764\,Mon recently as a `bona fide' RR Lyrae among the binary cepheid 
and RR Lyrae candidates of the Gaia DR2. 
On the basis of its proper motion anomaly value ($\Delta_{G2}=7.1$), which is the 
second highest of the Gaia RR Lyrae sample, the binary nature of the star is certain. Kervella et al. (2019a) 
estimated the upper limits of the semi-major axis ($a_{\mathrm{max}}\sim 300$~AU) and the 
orbital period ($P_{\mathrm{orb}}\sim 5000$~year) of this binary system.

With our present observations, we aimed to construct a spectroscopic time series of complete phase coverage for this potentially interesting star.
Two secondary targets (T\,Sex, $V_{\mathrm{max}}=9.91$~mag,
and HY\,Com, $V_{\mathrm{max}}=10.33$~mag) were also chosen to best utilise our observation time. 
The results for T\,Sex are published elsewhere (Benk\H{o} et al. 2021). 
This paper presents the results on V764\,Mon and HY\,Com.

\MakeTable{rcrrc}{12.5cm}{Used photometric data}
{\hline
Star & Time &  $\langle \sigma \rangle$ & $n$ & Source\\
     &  dd.mm.year &    (mmag)     &      &   \\
\hline

V764\,Mon &  12.03.1990--03.03.1993         & 8     &           97 &  (1) \\
          &  07.01.2019--02.02.2019         & 0.05  &         1093 &  (2) \\         
HY\,Com   & 29.12.2002--11.07.2009  &    30    &           368  & (3) \\
          & 01.04.2012--27.05.2012  &    72    &            1411 & (4) \\
           & 09.03.2013--28.11.2018  &    20    &             649 &   (5) \\
\hline
\multicolumn{5}{p{10.5cm}}{(1) Hipparcos (ESA 1997); 
(2) TESS (Ricker et al. 2015);
(3) ASAS-3 (Pojma\'nski et al. 2005);
(4) Qatar Exoplanet Survey (Bramich et al. 2014);
(5) ASAS-SN Variable Stars II (Jayashinge et al. 2019)
}
}

\section{The case of V764\,Mon}

\subsection{Photometric data}

We collected and analysed the two space-born photometric time series data of V764\,Mon observed by Hipparcos and \textit{TESS}
satellites.
Time length, number of observed data points and their typical accuracy of these data sets are given in Table~1.
Some large-scale ground-based surveys also observed V764\,Mon  
(ASAS-3, Pojma\'nski 2002, 2003; ASAS-SN, Shappee et al. 2014, Jayasinghe et al. 2019; KELT, Pepper et al. 2007, 2012; ZTF, Bellm et al. 2019, Masci et al. 2019).
However, V764\,Mon is well above the saturation limit of all these
surveys.
In some cases the image of the saturated stars can still be corrected and acceptable light curve can be reconstructed (e.g. Kochanek et al. 2017).
Nevertheless, checking V764\,Mon data sets, 
we came to the conclusion that such correction is not possible for any of the survey data, therefore, these data sets were not used for analysis.

For bright stars 
the Hipparcos astrometric satellite provided also some photometric measurements during its mission (van Leeuwen et al. 1997). The Hipparcos observation of V764\,Mon has been made by
a wide non-standard pass-band $H_{\mathrm p}$. 

The photometric space telescope \textit{TESS} (Ricker et al. 2015) is measuring the sky with four 4k$\times$4k CCD cameras, each of them providing 24$\times24$~deg field of view. The mission consists of 27-day-long observing runs in which the 
camera positions are fixed and they scan a single 24$\times$90~deg strip, called a `sector'. 
Thirteen such sectors almost cover an entire ecliptic hemisphere. 
The observations are continuous within the sectors, apart from some technical interruptions. One image is taken every 2 minutes by each camera.
Fifteen consecutive images are combined on-board into so-called full frame images (FFI)\footnote{FFIs are publicly available at the 
Mikulski Archive for Space Telescopes:
\url{https://archive.stsci.edu/}} which results in time-series observations of $\sim$30~min cadence. 
The spectral response function of the \textit{TESS} CCD cameras (Ricker et al. 2015) defines a broad ($\sim 4000$~\text{\AA}) 
non-standard band, most sensitive at red optical wavelengths, somewhat similar to the Johnson-Cousins band $I$.

V764\,Mon was so far observed by \textit{TESS} only in Sector 7.
We extracted the data on our target and its surroundings from the FFIs 
and processed the resulting small images by using a shell script written
by A. P\'al (Konkoly Observatory). The script relies on the tools of the {\sc{fitsh}} package (P\'al 2012) and beyond the target selection,
it performs a full image subtracting photometry, and also prepares the light curve for analysis.
The detailed description of the process is given in Plachy et al. (2021).
Each observing run of \textit{TESS} includes two complete orbits of the satellite around Earth. 
As a consequence, generally two different systematic trends can be seen in the raw light curves,
one for each circulation. This is the situation for V764\,Mon data as well. 
For eliminating these trends, quadratic functions were fitted to the two data parts separately, then these functions were subtracted from the respective sections\footnote{The used \textit{TESS} data are available here: {\url{https://konkoly.hu/KIK/data_en.html}}.}.

\subsection{Analysis of the photometric time series}

\begin{figure}[htb]
\includegraphics[angle=270,width=4.3cm]{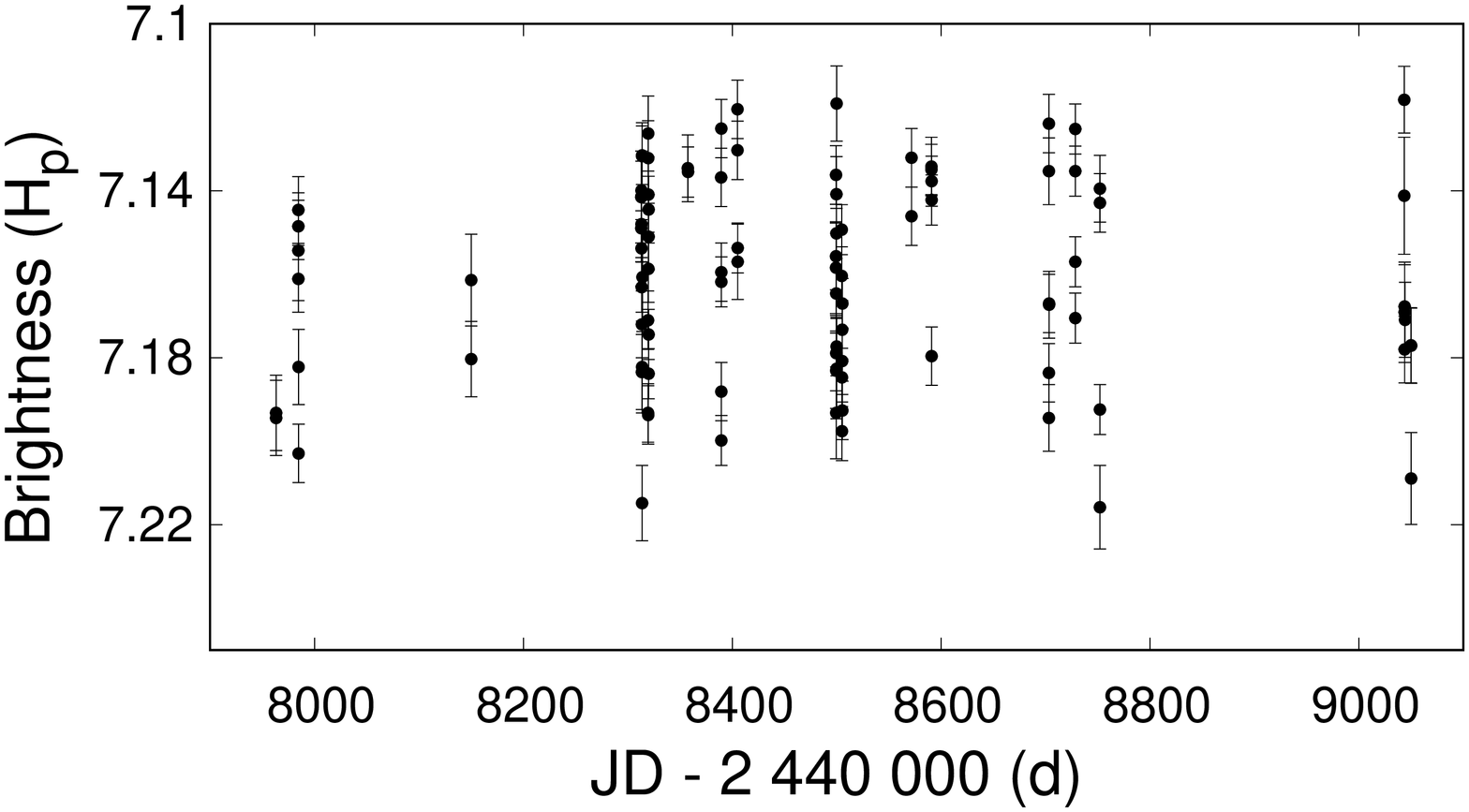}
\includegraphics[angle=270,width=3.88cm]{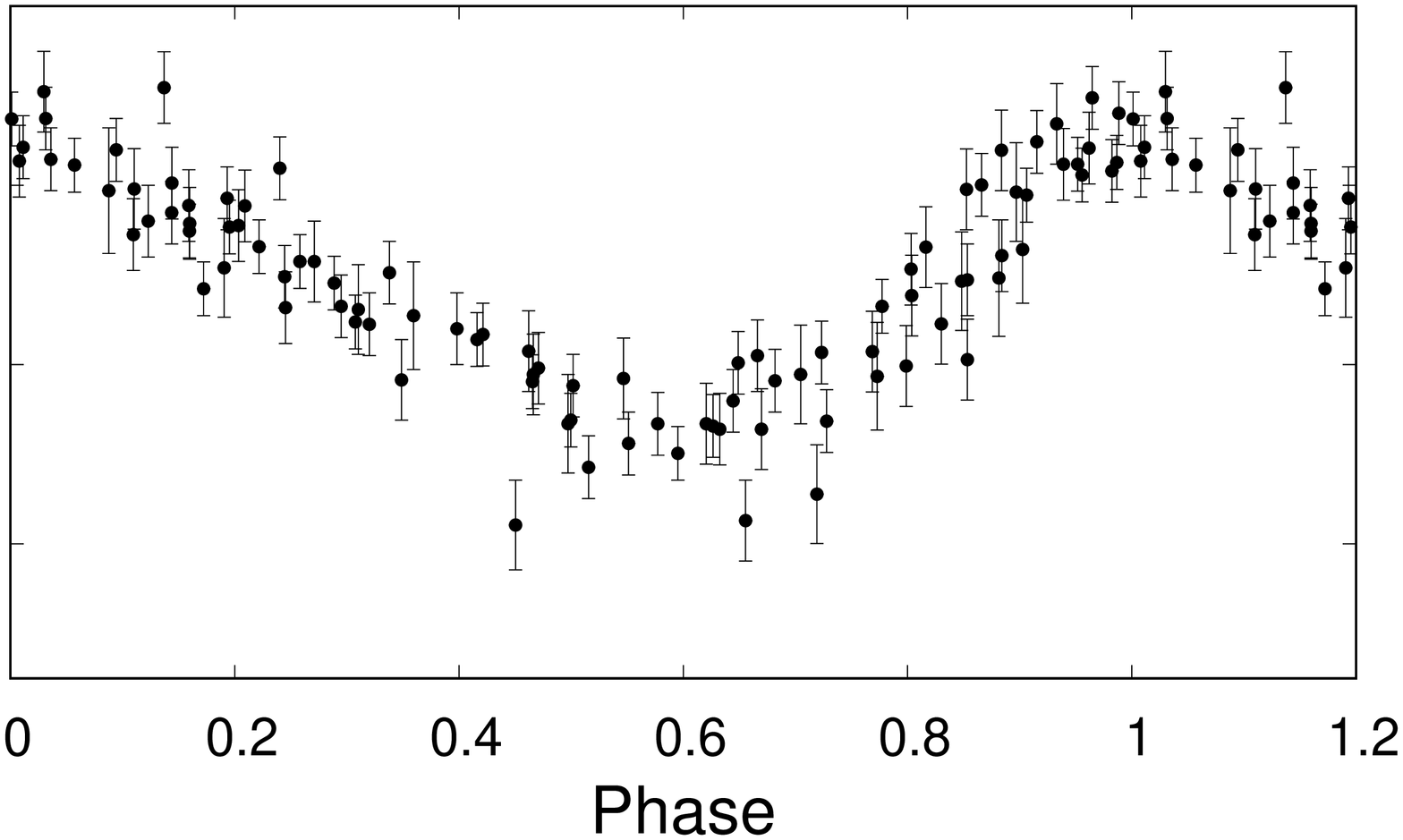}
\includegraphics[angle=270,width=4.3cm]{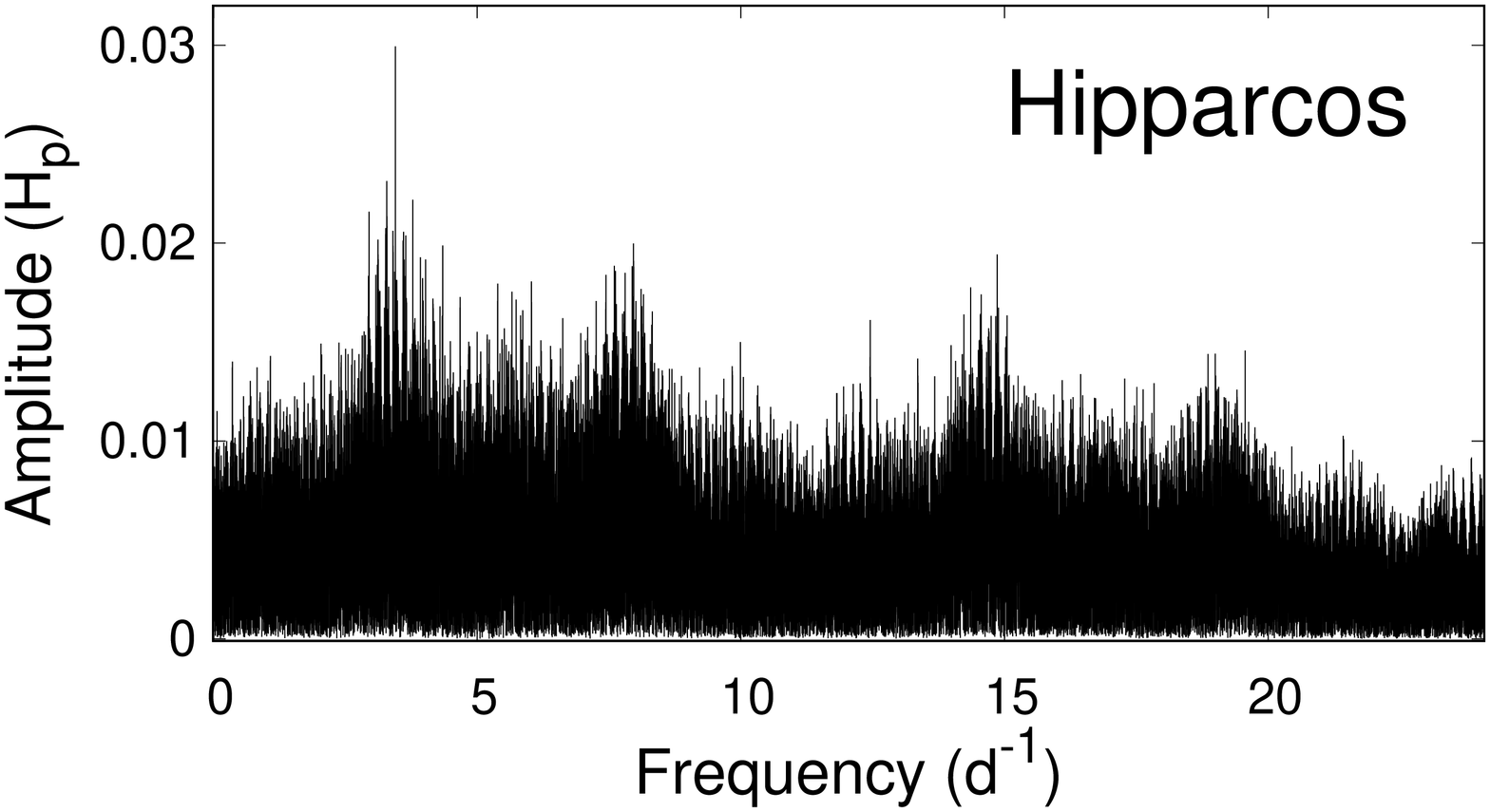}\\
\includegraphics[angle=270,width=4.3cm]{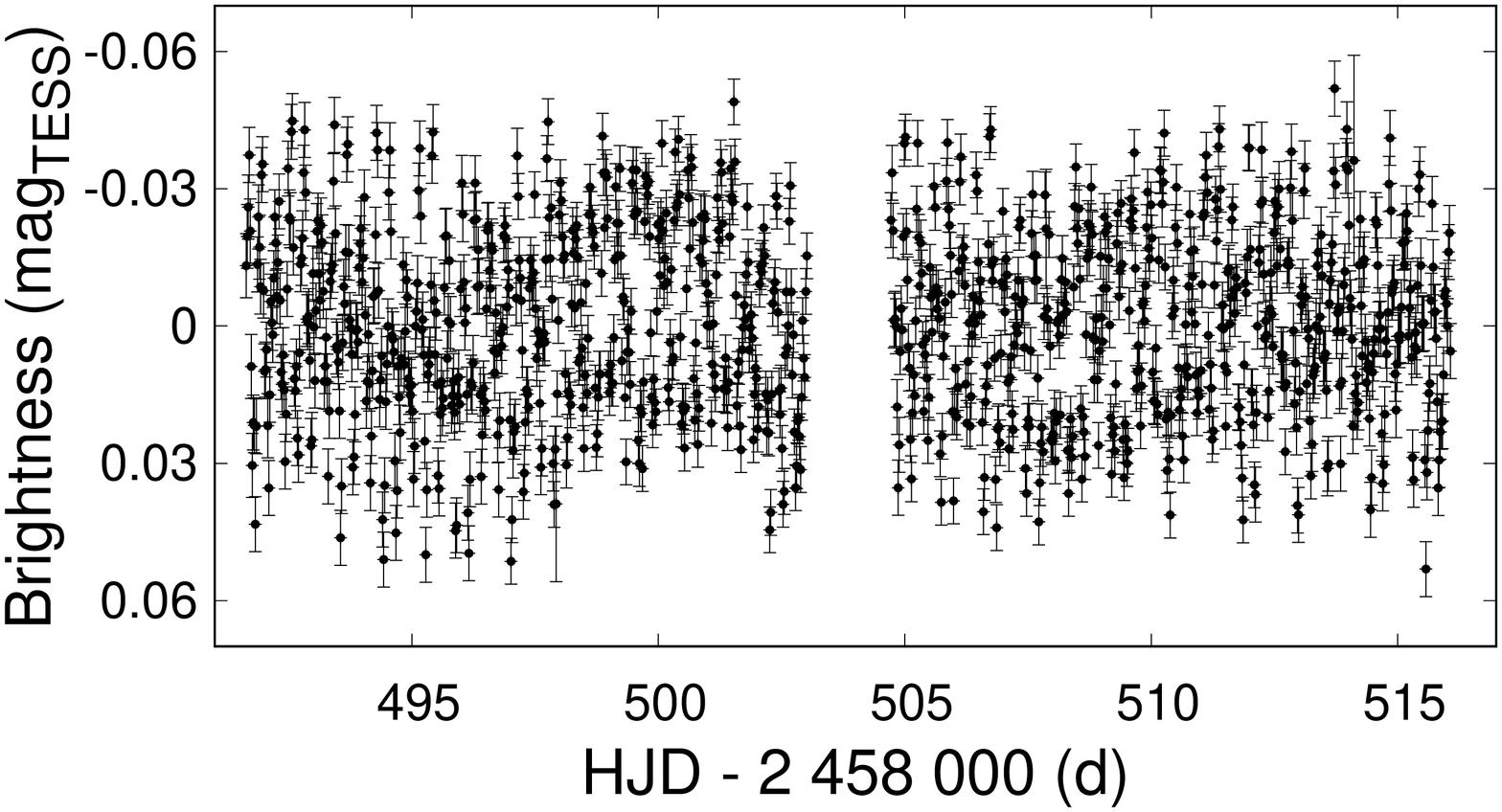}
\includegraphics[angle=270,width=3.88cm]{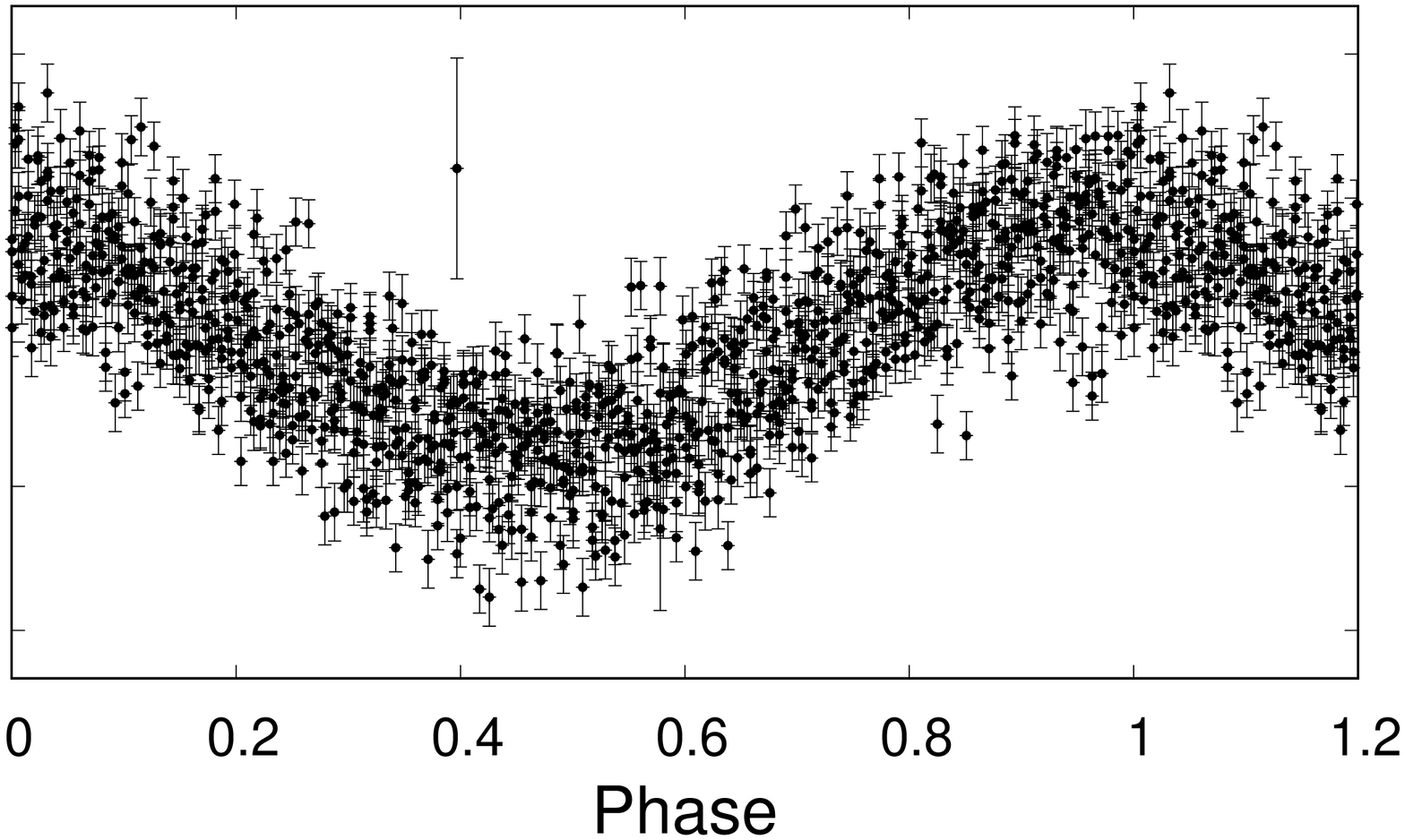}
\includegraphics[angle=270,width=4.3cm]{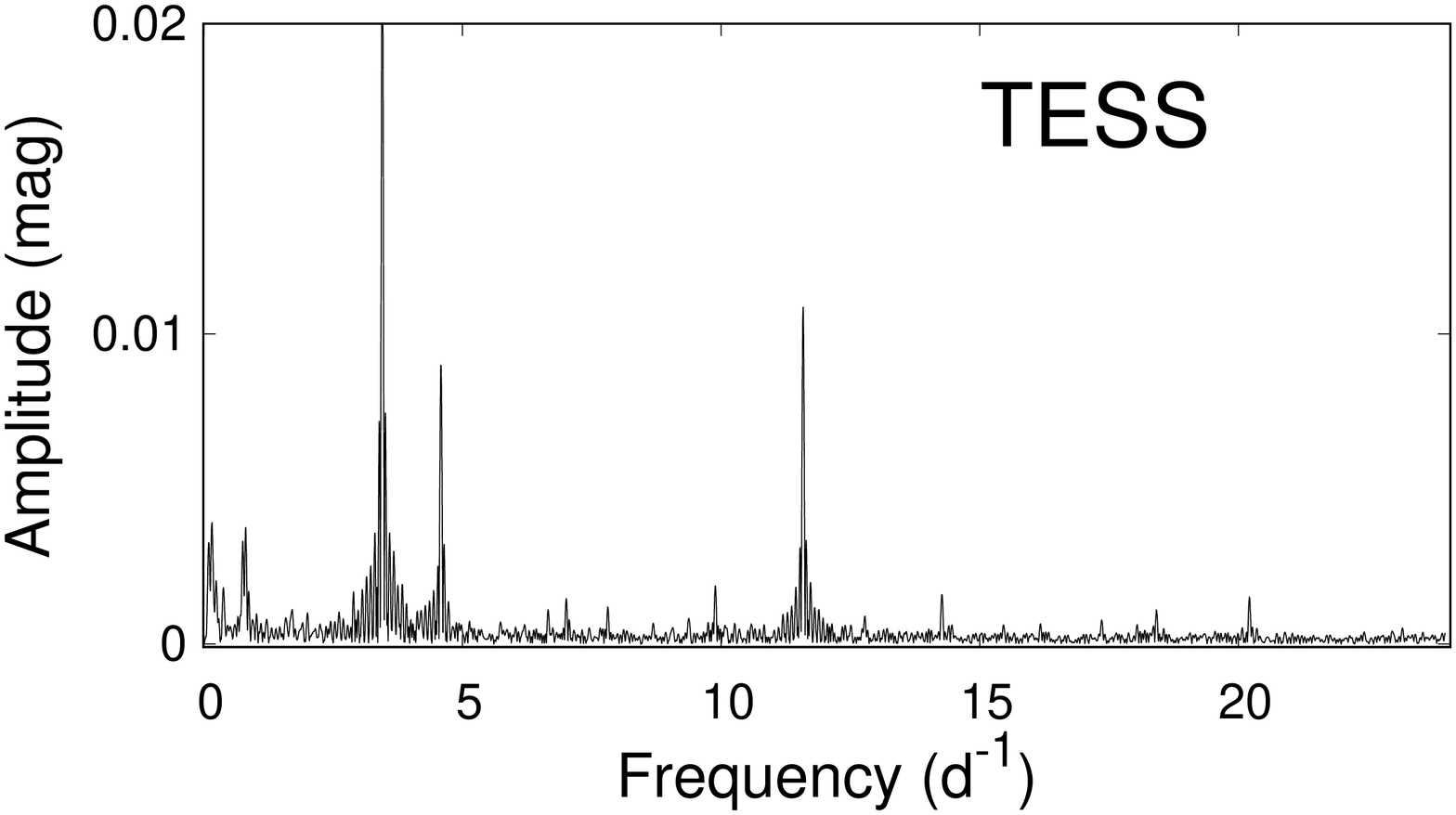}

\hspace{1cm}
\FigCap{
Photometric analysis of V764\,Mon. Observed light curves (left panels); phased light curves with the main period $P_1=0.290002$~d (middle panels); and the Fourier spectra of the corresponding data sets (right panels). First row shows Hipparcos data while second one shows the \textit{TESS} measurements.
}
\end{figure}

The time series analysis was performed by using the
tools of the {\sc{Period04}} software package (Lenz and Breger 2005).
The Fourier spectrum of the Hipparcos data show only one significant ($S/N=4.5$) 
peak at the frequency of $f_1=3.44825\pm0.00002$~d$^{-1}$ (top right panel in Fig.~1),
which is the known main period $P_1=0.290002\pm0.000002$~d. This period could really be a pulsation 
period of an RRc star, but the phased light curve (top middle panel in Fig.~1)
show an order of magnitude lower amplitude ($\sim 0.03$~mag) than expected for a typical RRc star. 
Although the Fourier spectrum shows some structure,
after we pre-whiten the data with the main frequency, no further significant 
($S/N>4$, Breger et al. 1993) peak appears in the residual spectrum. 

\MakeTable{lrccccrl}{10.5cm}{Significant frequencies of V764\,Mon from the \textit{TESS} data.}
{\hline
ID & Frequency   & $\sigma(f)$ &Amplitude  & Phase & $\sigma(\phi)$ & $S/N$ & ident.\\
   &  (d$^{-1}$) & (d$^{-1}$)  &  (mag)     &   (rad)   & (rad) &  &\\
\hline
$f_1$   &       3.44757  & 0.0004       &       0.0233       &  2.332 & 0.016 &47.57 &\\
$f_2$   &      11.58466  & 0.0008       & 0.0109        &       3.454 & 0.035 &25.96 & \\
$f_3$   &       4.58517  & 0.0010    & 0.0089      &            0.096 & 0.043 &19.00 &\\
$f_4$   &       0.15345  & 0.0023    & 0.0036       &           1.316 & 0.101 &4.82 &\\
$f_5$   &       0.80614  & 0.0022      & 0.0031      &          1.779 & 0.097 &4.70 &\\
$f_6$   &       9.88850  & 0.0067     & 0.0017      &           5.552 & 0.299 & 7.18&\\
$f_7$   &      14.26846  & 0.0085     & 0.0015      &           2.065 & 0.377 &7.75&\\
$f_8$   &       7.00563  & 0.0064    & 0.0014      &            5.521 & 0.284 &5.35 & $=f_2-f_3$\\
$f_9$   &      20.21424  & 0.0067     & 0.0014       &          4.538 & 0.295 &8.37 &\\
$f_{10}$  &     7.81177  & 0.0081     & 0.0012      &           0.960 & 0.357 &5.57 &$=f_5+f_8$\\
$f_{11}$  &    18.41987  & 0.0050    & 0.0011      &            5.495 & 0.221 &6.27 &\\
$f_{12}$  &     6.65780  & 0.0076    & 0.0010     &             3.701 & 0.335 &4.08  &\\
\hline
\multicolumn{8}{p{10.5cm}}{
}
}
The Fourier spectrum of the \textit{TESS} data shows several highly significant frequencies. The highest-amplitude peak is at the main frequency ($f_1=3.44757$~d$^{-1}$, bottom right panel in Fig.~1). Folding the light curve with this period, we obtain a thick phase curve (middle bottom panel in Fig.~1), indicating the multi-periodic nature of the signal.
Two high peaks are situated at $f_2=11.58466$~d$^{-1}$ and at  $f_3=4.58517$~d$^{-1}$ frequencies, and some lower-amplitude peaks 
are also detectable. Yet, no evident harmonics of $f_1$ can be seen. 

Traditionally, the frequency content of multi-periodic stars have been determined by successive pre-whitening 
steps starting with the highest-amplitude frequency, and continuing with the next highest-amplitude frequency of
the residual, and so on. As Balona (2014b) has demonstrated,
for many low-amplitude frequencies, which are common with stars measured by space photometry, 
this method could give misleading results. He suggested no or as few pre-whitening steps as possible. Following this, we pre-whiten the data with the three highest amplitude frequencies ($f_1$, $f_2$ and $f_3$). The residual spectrum shows further nine significant frequencies (see Table~2).  Three of these ($f_4$, $f_5$ and $f_{11}$) are significant in the original spectrum as well. 
After pre-whitening the original data with all 12 frequencies, no further significant frequencies were found in the residual.
Examining the obtained frequencies, 
we find two linear combination frequencies ($f_8$ and $f_{10}$), that is, ten independent 
frequencies describe the light curve within the observed accuracy.

The result of the frequency analysis is presented in Table~2. 
The frequencies are arranged in order of decreasing amplitude.
After the designation of the frequency (col.~1) we show
the frequency values and their 1$\sigma$ errors (cols~2--3), the amplitude (col.~4),
the phase with their errors (cols~5--6), signal-to-noise ratio (col.~7) and 
possible identification (col~8). The uncertainties of Fourier parameters are estimated by the 
analytical formulae of Breger et al. (1999) implemented in the {\sc{Period04}} package.  
These formulae uniformly give an error of 0.0004~mag for the amplitudes, which have not been indicated individually.
We mention that the two largest-amplitude non-significant peaks in the spectrum of the Hipparcos data can be identified with $f_{10}$ and $f_7$, respectively. 
This suggests that 
the amplitudes of the pulsations are changing in time.

Based on the \textit{TESS} spectrum, the RRc classification can no longer be maintained.
The small-amplitude multi-periodic light-curve variation together with its spectral classification (A5IV)
suggests that V764\,Mon is a possible $\delta$ Scuti star.
Or, since its ten independent frequencies are distributed between 0.15345 and 20.2142~d$^{-1}$,
maybe a hybrid $\delta$ Scuti/$\gamma$ Dor star.
This classification is controversial as to whether it says more than the $\delta$ Scuti type alone because Balona (2014a) and Balona et al. (2015) have shown that low-frequencies associated with the g-modes and characteristic of $\gamma$ Dor stars can be detected in almost all \textit{Kepler} $\delta$ Scuti stars.
The frequency distribution of V764\,Mon, shown in the bottom right panel of Fig.~1, is similar 
to the \textit{Kepler} $\delta$ Scuti stars described by Balona and Dziembowski (2011), as stars with 
one dominant mode with a side frequency. The mode density of V764\,Mon (2.0~per d$^{-1}$) is also typical in this sub-group.
Summarising our photometric analysis, it can be concluded that 
V764\,Mon is a low-amplitude multi-periodically pulsating star, probably 
a $\delta$~Scuti or related variable.

\subsection{Spectroscopic data}

Spectroscopic time series were obtained with the ACE echelle spectrograph mounted on the 1-m RCC telescope
set up in Piszk\'estet\H{o} Mountain Station of the Konkoly Observatory.
The spectra cover the 4150\,--\,9150~\text{\AA} wavelength range with a resolution of $R\approx$20\,000.
The log of the spectroscopic observations are given in Table~3.
The integration time was set to 30~min which provides signal-to-noise ratios between 85 and 380,
depending on the weather conditions
(see column 4 in Table~3 for averaged nightly $S/N$ values). These S/N values were estimated by using the tool 
of {\sc{iSpec}} package (Blanco-Cuaresma et al. 2014; Blanco-Cuaresma 2019).

The spectra were processed by using standard {\sc IRAF} (Tody 1986, 1993) tasks including bias,
dark and flat-field corrections, aperture extraction, and wavelength calibration. For the latter purpose we used
thorium-argon calibration images which were taken after every three object frames.
The normalisation,  cosmic-ray  filtering and order  merging were performed by our
 {\sc python} scripts.  Each spectrum was also corrected to the barycentric frame\footnote{
The final normalised spectra are available from the authors upon request.}.

\MakeTable{rccrr}{12.5cm}{Log of the spectroscopic observations}
{\hline
Star & Night & JD &  $\langle S/N \rangle$  & $n$ \\
     & (yyyy-mm-dd) &  (-2 400 000)  &       &         \\
\hline
V764\,Mon & 2015-03-05   &   57087        & 84    &           3  \\
          & 2015-03-06   &   57088        & 163   &           10  \\
          & 2015-03-09   &   57091        & 376   &           8  \\
HY\,Com   & 2015-03-06   &   57088        & 18    &           8  \\
          & 2015-03-08   &   57090        & 43    &           12  \\
          & 2015-03-09   &   57091        & 27    &           6  \\
\hline
\multicolumn{5}{p{7cm}}{
Columns contain: name of the target star; night of the observation; Julian day; average signal-to-noise ratio of the single exposures; number of spectra}
}

\subsection{Spectroscopic results}

Radial velocities were calculated by cross-correlating the spectra with a metallic line mask containing 622 metallic lines 
between 4800 and 5600~\text{\AA} from a synthetic A-type spectrum. We fitted a rotation-broadened line profile to the time-averaged mean pulsation profile for determining radial velocity $v_{\mathrm{rad}}$ and $v\sin i$ simultaneously, using an in-house developed Python application relying on packages of {\sc Phoebe 2.0} (Pr{\v{s}}a et al. 2016). The fitting process also provided micro- and macro-turbulences, however, the results on the turbulences were too uncertain and degenerate, while their values did not affect $v_{\mathrm{rad}}$ at all, and also affected $v\sin i$ only marginally. The uncertainties of the derived turbulences are caused partially by the limits of our data, but also the non-radial, time- and location-dependent pulsation motions.
The  systematic  errors resulted in the data processing and the stability of the wavelength 
calibration system of  the ACE instrument  are  better than 
0.36 km\,s$^{-1}$, based on observations of radial velocity standards (Derekas et al. 2017). 
The radial velocities and their standard deviations are given in Table~4.
The average radial velocity is $v_{\mathrm{rad}}=29\pm 1$~km\,s$^{-1}$ while 
$v\sin i=188\pm 5$~km\,s$^{-1}$.
The latter value is
an order of magnitude higher than the known values of RRc stars
($\sim 2-15$~km\,s$^{-1}$, Preston et al. 2019), while typical for near-main-sequence A-F stars,  including $\delta$ Scuti-type pulsators.

The spectral measurements were optimised with the assumption that V764\,Mon is a bright RRc star.
Once this was disproven, the observations were also found to be sub-optimal for the study of
a potential $\delta$~Scuti pulsator. E.g., preparing a proper radial velocity curve of a $\delta$~Scuti star needs an order of higher precision 
than ours, on one hand because 
the expected amplitude of the radial velocity curve is 
far less than that of an RRc (of about 5~km\,s$^{-1}$, see e.g. Yang and Walker 1986; Mathias and Aerts 1996), and on the other hand because the uncertainties of the radial-velocity measurements of fast rotators are significantly higher. 
Nevertheless, if we fold the radial velocity data of V764\,Mon with its dominant period $P_1$ (left panel in Fig.~2),
the curve show a definite phase-dependence,
but it does not allow a detailed examination, because of the relatively large errors, and because of the relatively few number of data points, considering the multi-periodic nature of the pulsation.

\MakeTable{cccccc}{7cm}{Radial velocities of V764\,Mon}
{\hline
BJD &  $v_{\mathrm{rad}}$  & $\sigma(v_{\mathrm{rad}})$ & BJD &  $v_{\mathrm{rad}}$  & $\sigma(v_{\mathrm{rad}})$ \\
    (d)  & km\,s$^{-1}$ &  km\,s$^{-1}$        &    (d)  & km\,s$^{-1}$ &  km\,s$^{-1}$        \\
\hline
2457087.38387  &  36.12  &   1.49 &  2457088.42919   &  26.31 &   0.72\\
2457087.40482  &  34.11  &   1.53 &   2457088.45018  &  27.68 &   0.89\\
2457087.42577  &  40.69  &   2.32 &   2457091.27534  &  29.10 &   0.32\\
2457088.25168  &  30.69  &   0.46 &   2457091.29628  &  28.99 &   0.43\\
2457088.27263  &  30.19  &   0.69 &   2457091.31723  &  27.67 &   0.50\\
2457088.29362  &  28.08  &   0.58 &    2457091.34924 &  28.78 &   0.37\\
2457088.31748  &  28.14  &   0.45 &   2457091.37021  &  29.34 &   0.37\\
2457088.33847  &  27.67  &   0.47 &   2457091.39116  &  29.56 &   0.49\\
2457088.36434  &  28.76  &   0.56 &    2457091.42026 &  30.41 &   0.33\\
2457088.38530  &  28.25  &   0.47 &    2457091.44123 &  30.20 &   0.48\\
2457088.40628  &  34.06  &   0.62 & & & \\
\hline
\multicolumn{6}{p{7cm}}{
}\label{tab:V764_vrad}
}

\begin{figure}[htb]
\includegraphics[angle=0,width=6.3cm, trim=0cm 0 0cm 0]{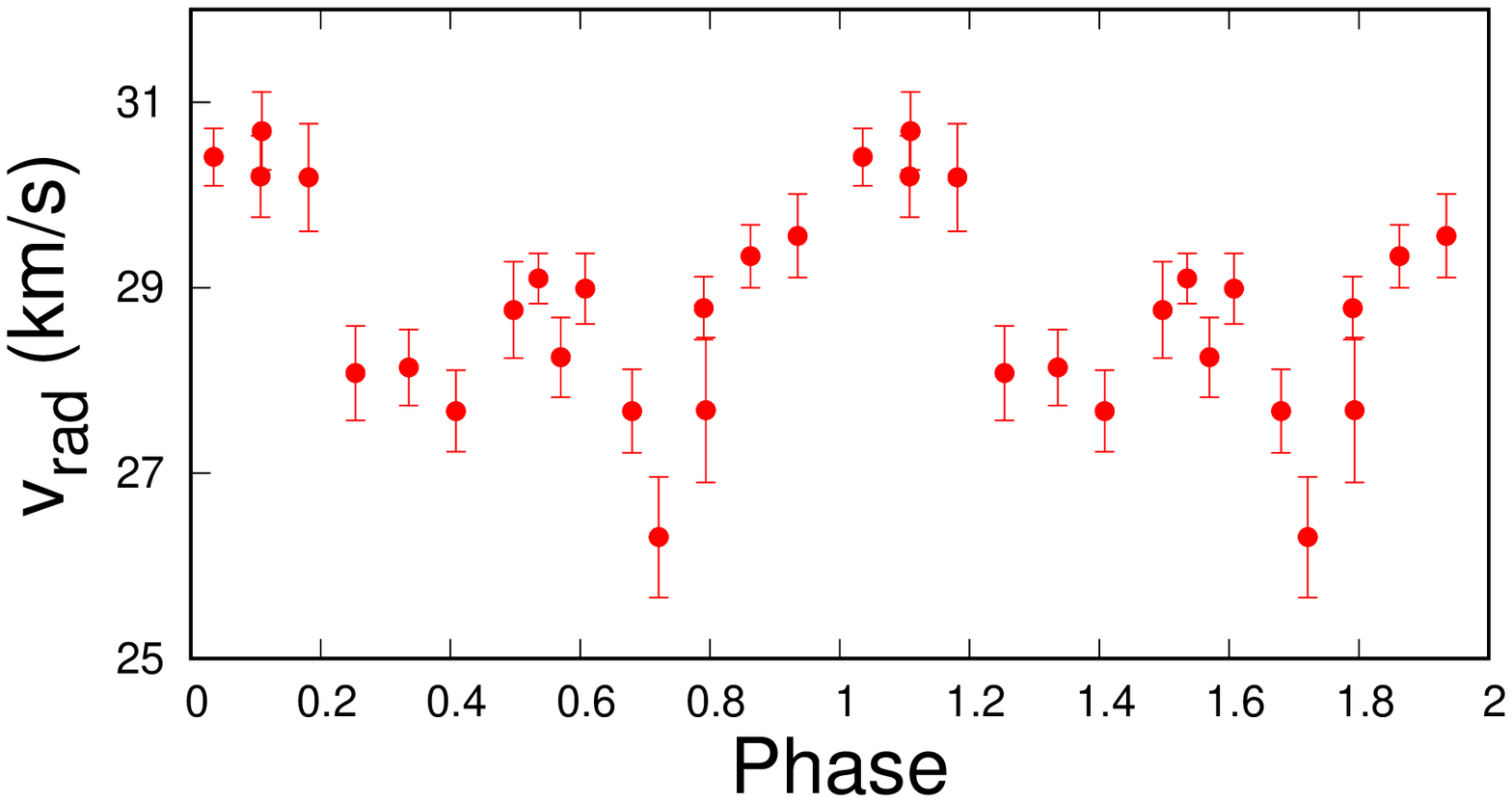}
\includegraphics[angle=0,width=6.3cm, trim=0cm 0 0cm 0]{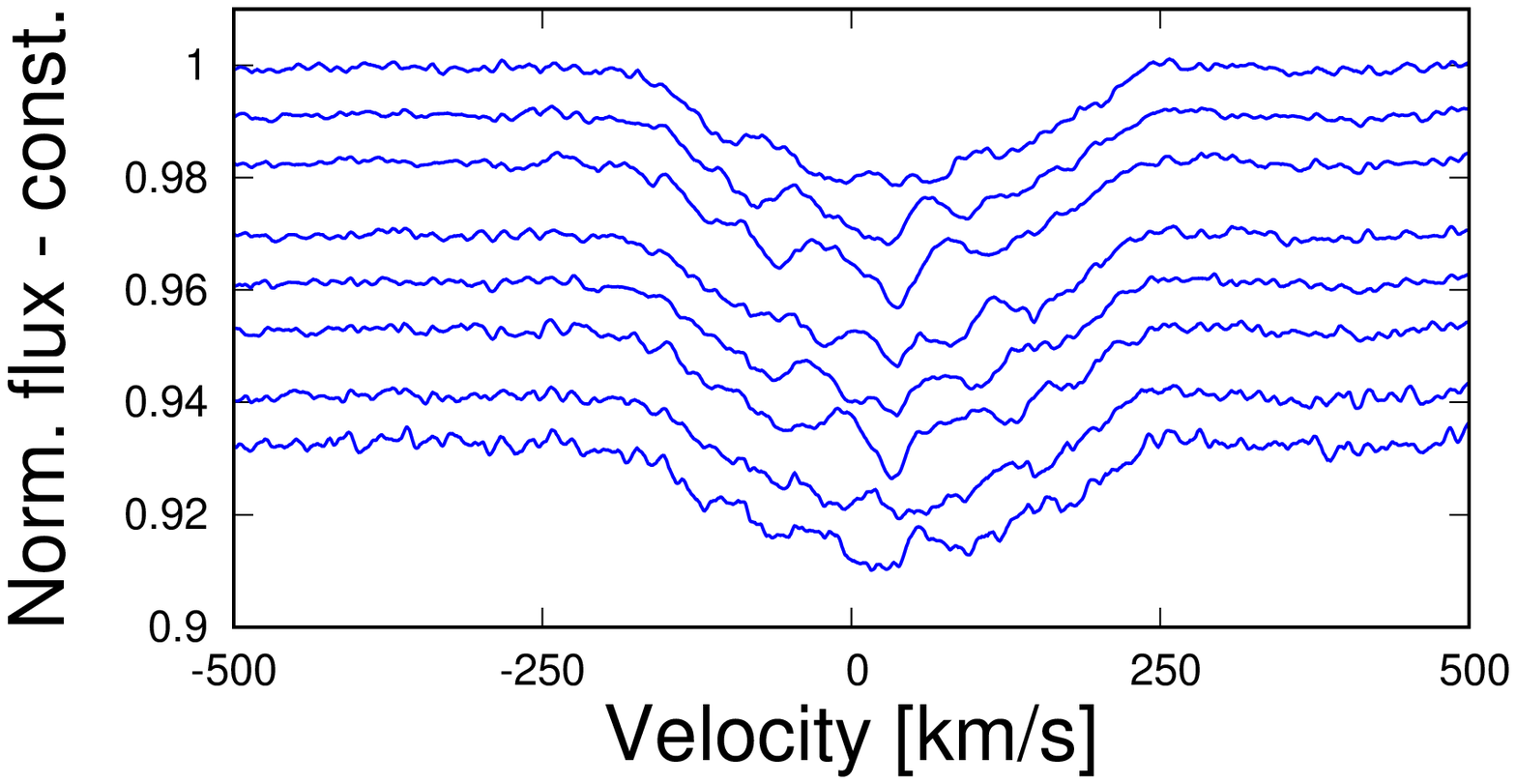}\\
\FigCap{On the left:
Phased radial velocities of V764\,Mon with its dominant
period $P_1=0.29$~d. The scatter in radial velocity is partly attributed to the multi-periodic nature of the pulsation. 
On the right:
Line profile variation of V764\,Mon in the night of 2015-03-09.
Consecutive profiles are plotted from top to bottom with vertical offsets from each other. 
The offsets are proportional to the differences in the observation time of each spectrum.
Red-ward propagating ripples in the profiles can be observed, typical of $l \ge 1$ azimuthal-order, non-radial pulsations.
}
\end{figure}

Investigating individual cross-correlation functions two phenomena can be seen: 
(i) The line profiles show red-ward propagating distortions caused by non-radial pulsation.
The time series spectroscopy of $\delta$~Scuti stars is primarily used to identify their non-radial modes. 
However, for a successful mode identification, each of the known methods
(see for reviews e.g. Mantegazza 2000, Zima 2008)
requires at least an order of magnitude more individual spectra, preferably with the resolution of at least 40\,000 and $S/N>200$. 
In the best night (2015-03-09) the $S/N$ ratio is above this limit and the line profiles variations can clearly be followed (right panel in Fig.~2).

(ii)  By combining all spectra, the line-profile variations caused by the pulsation are mostly averaged out, while a clearly visible absorption feature 
emerges at around 40 km\,s$^{-1}$(Fig.~3, left panel). 
This signal most probably belongs to the companion star of this binary system. 
The spectra are unsuitable for decomposition. This might require better $S/N$ ratio.
We can, however, fit the average radial velocity of the companion (Fig.~3 in the right) 
as $v_{\mathrm{rad}}=34.7\pm 2$~km\,s$^{-1}$ and the projected rotation velocity $v\sin{i} = 12\pm 3$~km\,s$^{-1}$. 
If the rotation axes are parallel, the latter would mean slow rotation, but in the case of wide binaries, 
the rotation axes are not necessarily aligned.
The equivalent width of the cross-correlated line profile of the companion is $\sim 16$\% of that of the V764\,Mon,
that is, the secondary star is certainly significantly fainter than the  $\delta$~Scuti pulsator primary, but a quantitative estimate is not possible without knowing the spectral types of both components.
Without this information, we cannot estimate the luminosity ratio. A plausible assumption is, however, that the companion is a red main-sequence star.
\begin{figure}[htb]
\includegraphics[angle=0,width=6.5cm, trim=0 0 0.5cm 0]{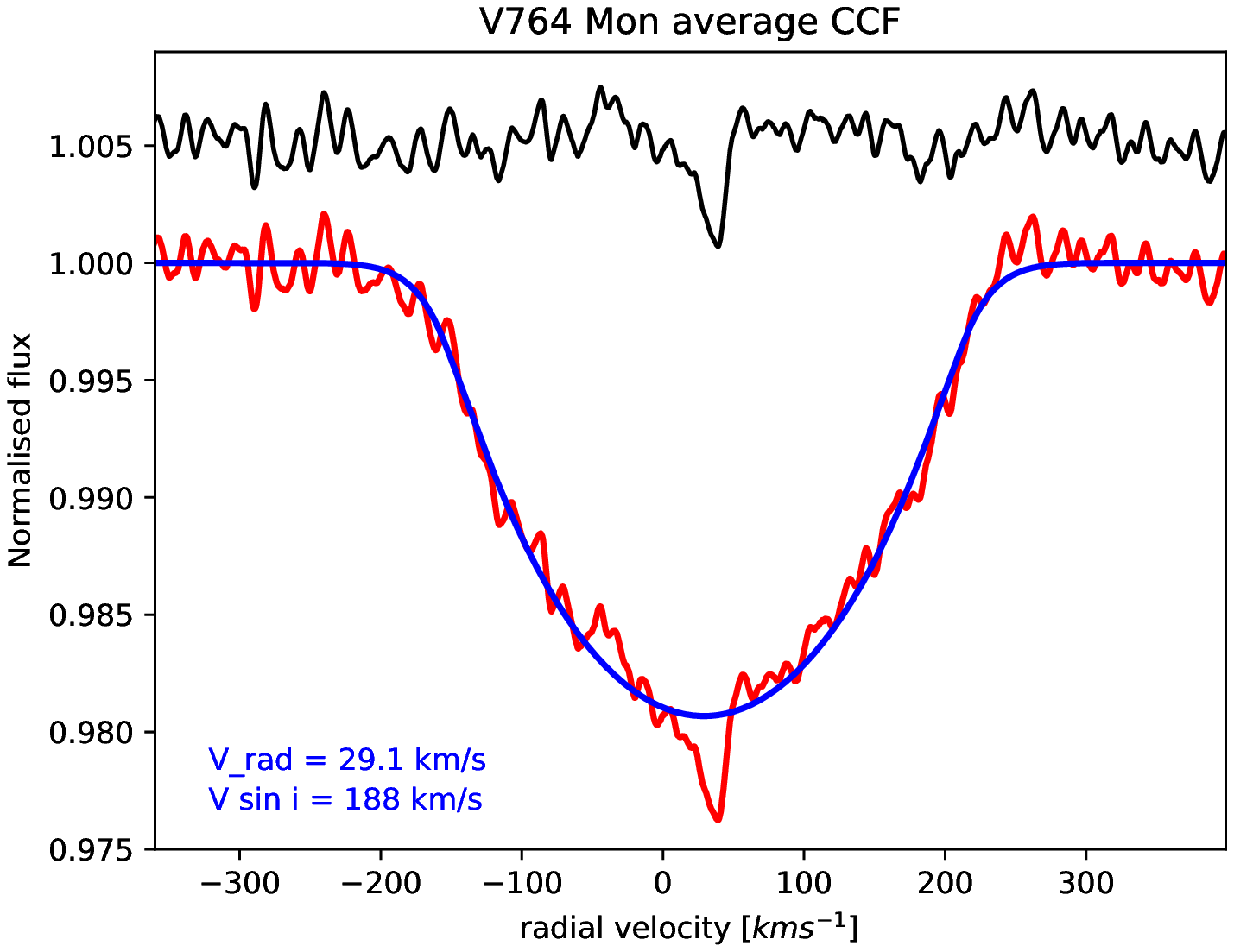}
\includegraphics[angle=0,width=6.5cm, trim=0.5cm 0 0cm 0]{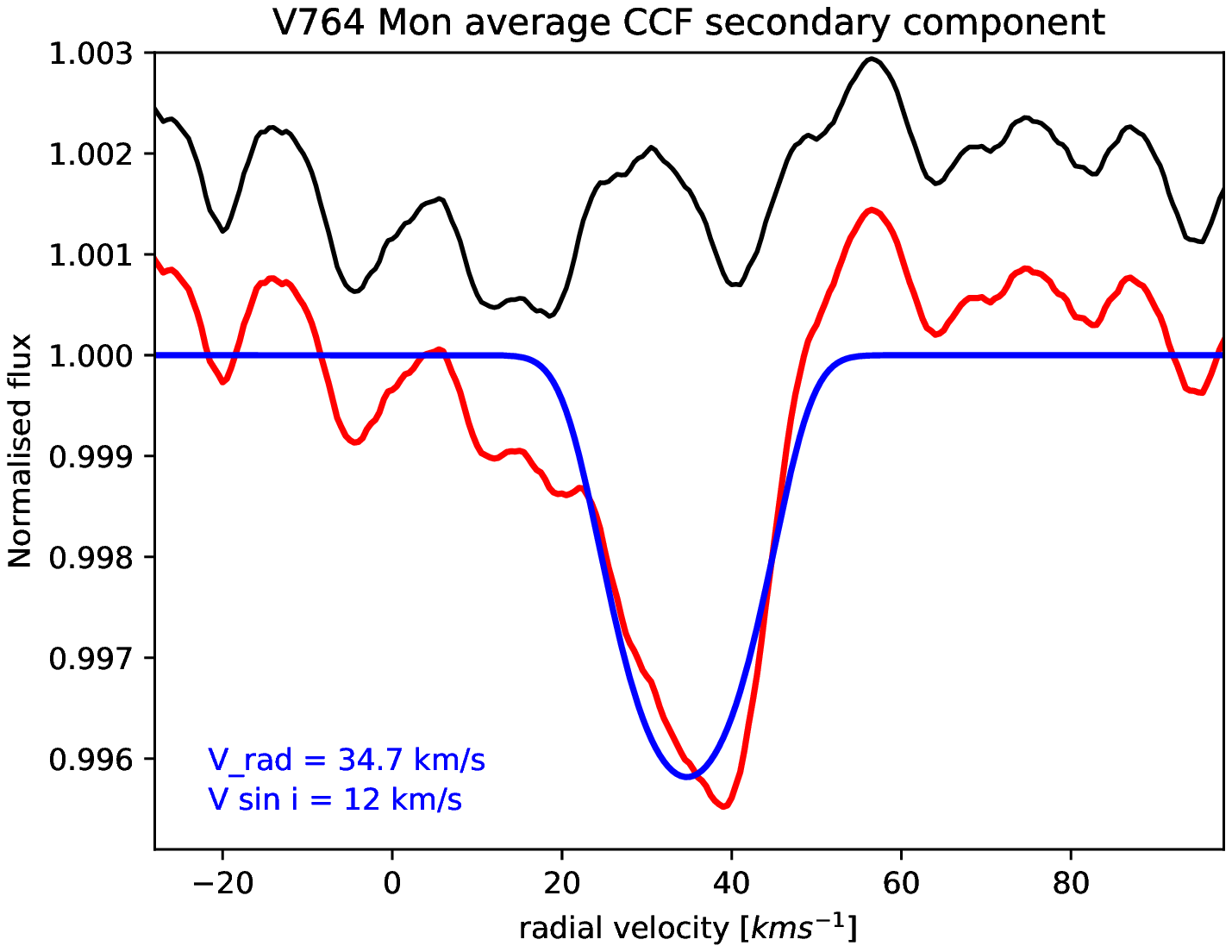}\\
\FigCap{In the left:
Average of the weighted cross-correlation function of the spectroscopic time series of V764\,Mon (red line)
and the fitted line profile (blue). The black line shows the
averaged profile after we subtracted the fitted profile from it and shifted for clarity (black).
In the right: The fit of the secondary component's line profile using the subtracted average (top curve in the left panel). 
Line colours have the same meaning as in the left panel of the figure.
}
\end{figure}

From 60 to 80 percent of the stars in the Galactic field are members of a binary or multiple system. 
Although the presently known ratio for $\delta$~Scuti stars is smaller than this, lots of binary systems are known
(Liakos and Niarchos, 2017, Lampens et al. 2018).
Therefore, it is not surprising that V764\,Mon is member of a binary system, 
in contrast with the binary RR Lyrae scenario suggested by Kervella et al. (2019a, b).

\subsection{Rotation}

Although V764\,Mon is not listed in the $\delta$~Scuti catalogue of Rodr\'{\i}guez et al. (2000), 
Bush and Hintz (2008) included it in their $\delta$~Scuti rotational velocity survey 
and they have determined a projected velocity of $v\sin{i}=131\pm 8$~km\,s$^{-1}$. 
As we have seen, 
our projected rotation velocity is significantly higher than the measurement of Bush and Hintz (2008).
The phase dependent radial velocity due to
pulsation could cause some
difference in the $v\sin i$ measurements, but this effect cannot explain the whole difference. 
Bush and Hintz (2008) used a standard technique for determining $v\sin i$ which imply the assumption of a nearly linear 
relation between the line width and $v\sin i$. 
This method could, however, systematically underestimate the projected equatorial velocities,
especially for rigidly rotating star with radiative envelope,
if the gravity darkening is not taken into account properly (see e.g. Townsend et al. 2004).
The rotation study of A--F stars by Ammler-von Eiff and Reiners (2012) demonstrated well  
that the hotter ($>7000$~K) and faster rotating ($v\sin i> 50$~km\,s$^{-1}$) stars typically rotate as rigid bodies.
V764\,Mon fulfils both conditions, it is probably a rigid rotator and
the effect of the gravity darkening explains the difference between $v\sin i$ measurements.
Since a very large fraction of spectroscopic binaries are members of a triple or multiple system (Tokovinin et al. 2006),
there might be an alternative explanation. 
If the V764 Mon binary has a third component on a wide orbit
its inclination could change in the time between the two measurements (2002\,--\,2003 and 2015)
due to an orbital plane precession caused by the third star.

\begin{figure}[htb]
\includegraphics[angle=270,width=9.5cm, trim=0cm 0 0cm 0]{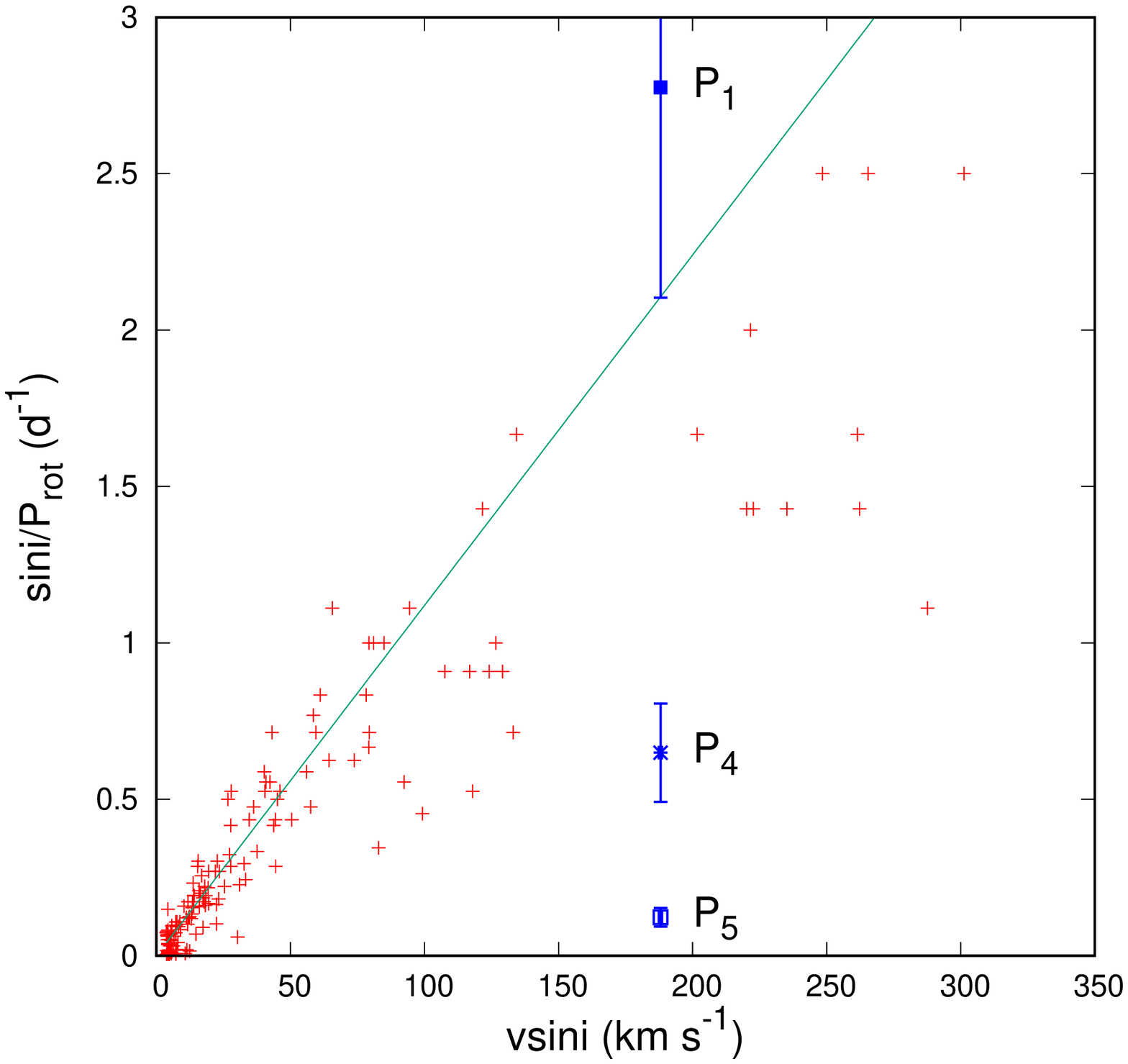}\\
\FigCap{
Correlation between projected velocity $v\sin i$ and $\sin i/P_{\mathrm{rot}}$
for A-F stars determined by Ammler-von Eiff and Reimers (2012). Our linear fit for 
the upper envelope is shown with the continuous line.
The three vertical bars mark the positions of the found
three potential rotation periods of V764\,Mon.
The vertical length of the bars result from the allowed $\sin i$ ranges.
}
\end{figure}
We attempted to identify any \textit{TESS} frequency that might correspond to the rotation. 
When we plot the projected velocity $v_{\mathrm t}\sin i$ versus $\sin i /P_{\mathrm {rot}}$
determined by Ammler-von Eiff and Reimers (2012), we obtain a 
clear correlation (Fig.~4).
It naturally follows from the relationship between the rotation period $P_{\mathrm {rot}}$ 
and the tangential velocity $v_{\mathrm t}$:
\begin{equation}
    v_{\mathrm t} = \frac{2\pi R }{P_{\mathrm {rot}}}. 
\end{equation}
If we fit the upper envelope in Fig.~4 as
\begin{equation}
      \frac{\sin i}{P_{\mathrm {rot}}} = a v_{\mathrm t}\sin i, 
\end{equation}
we obtain $a=0.0112\pm 0.00027$ 
which gives a lower limit for the rotation period: 
$a v\sin i=2.1$, $P_{\mathrm {rot}} > \sin i/2.1$~d.

Using the formulae 
for the critical rotation speed $v_{\mathrm {crit}}$ (e.g. Towsend et al. 2004):
\begin{equation}
    v_{\mathrm {crit}}=\sqrt{G \frac{M}{R_{\mathrm e}}}=\sqrt{2G \frac{M}{3R_{\mathrm p}}},
\end{equation}
where $G$ is the gravitational constant, $M$ is the stellar mass,
$R_{\mathrm e}$ and $R_{\mathrm p}$ are the equatorial and polar radii,
with the parameters of V764\,Mon its critical speed is about
$v_{\mathrm {crit}}\sim$308~km\,s$^{-1}$. (Here we accepted the equatorial radius of Gaia DR2: $R_{\mathrm e}=4.52$~R$_{\odot}$).
Then the minimum value of the $\sin i$ can be around 0.61, and the lower limit of the rotation period is $P_{\mathrm {rot}} > 0.29$~d.

There are three detected periods in the \textit{TESS} data 
above this limit:
$P_1=0.290$, $P_4=1.240$ and $P_5=6.517$~d.
According to the above rough estimation, 
it is possible that $P_1$ is not the main pulsation period but a rotational variation. If so, the star is very close to the 
break up limit. Although it would not be completely unprecedented, since several known $\delta$~Scuti stars rotate as 
fast as about 80\,--\,90\% of their break-up limit (e.g. Su\'arez et al. 2005; Monnier et al. 2010), but not very 
likely, nevertheless (see also Fig.~4). Considering the distribution shown in Fig.~4, 
$P_4$ is the most probable 
value of the rotation period in
V764\,Mon. In this case, if we estimate the inclination using the formulae (1, 2) and the radius estimation ($R=4.52$~R$_{\odot}$), we find that $i$ is very close to 90 degrees.

\subsection{Physical parameters}

\MakeTable{rrcccr}{12.5cm}{Estimated physical parameters of V764\,Mon.}
{\hline
Parameter                & McDonald$^1$   & LAMOST$^2$                        & DR2$^3$             & EDR3$^4$ &  this work \\
\hline
$T_{\mathrm{eff}}$ (K)   & 7197$\pm 317$    & \textit{7234}$\pm$\textit{48}  & 7455$\pm{324}$       & 8063$\pm40$ & 8090$\pm160$\\
$\log g$                 &                  & \textit{3.71}$\pm$\textit{0.06}&                      & 3.4986$\pm0.1$ & 3.60$\pm0.1$ \\
$[$Fe/H$]$               &                  & \textit{0.146}$\pm$\textit{0.1}&                      & $-0.015\pm0.04$ & $-0.30\pm0.3$ \\
$L$ (L$_{\odot}$)        & 62.97            & \textit{26.8$\pm$6.3}          & 56.84$\pm8.5$       & \textit{83$\pm$29} & \textit{63$\pm$30} \\
$M$ (M$_{\odot}$)        &                  & \textit{2.0$\pm$0.1}           &                      & 2.5209 & \textit{2.26$\pm$0.65}\\
$R$ (R$_{\odot}$)        &  \textit{5.24}   & \textit{3.3$\pm$0.3}           & 4.52$\pm0.31$        & \textit{4.7$\pm$0.8}  & \textit{4.02$\pm$0.8} \\
\hline
\multicolumn{6}{p{10.5cm}}{
(1) McDonald et al. (2012);
(2) from the empirical relation of Qian et al. (2018);
(3) Gaia DR2 (Andrae et al. 2018);
(4) Gaia EDR3 (Gaia Collaboration 2020).
The normal fonts mean literature values while 
italic fonts are the calculated parameters.}
}

McDonald et al. (2012) estimated the effective temperature and luminosity 
of more than 100\,000 stars observed by the Hipparcos satellite comparing
their optical and infrared spectral energy distribution (SED) with model atmospheres. 
The obtained values of V764\,Mon are shown in column 2 in Table~5.
No infrared excess, i.e. circumstellar material was detected for V764\,Mon.

Qian et al (2018) published empirical relations for physical parameters of 
$\delta$~Scuti stars based on the LAMOST spectroscopic survey. 
They isolated a separate group of multi-periodic stars which are significantly cooler than 
the normal $\delta$~Scuti stars. They pulsate with 0.09 and 0.22~d periods, therefore, V764\,Mon 
cannot be a member of this group. Qian et al. (2018) consider longer period ($\sim 0.35$~d) stars as `normal' $\delta$~Scuti stars. 
Applying their formulas to V764\,Mon we obtain the value of column 3 in Table~5.
This effective temperature is in good agreement with that determined McDonald et al. (2012) from their SED fit. 

Gaia DR2 (Gaia Collaboration 2016, 2018, Andrae et al. 2018) estimated the effective temperature, luminosity and radius of stars 
on the basis of their three-band photometry ($G$, $G_{\mathrm{BP}}$ and $G_{\mathrm {RP}}$) and parallax (col. 4 in Table~5).
The recently released Gaia EDR3 (Gaia Collaboration 2020) contains updated and several additional parameters for V764\,Mon
($\log g$, [Fe/H] and $M$) that were not yet included in DR2 catalogue (column 5 in Table~5), even though these values are actually based on DR2 measurements (Recio-Blaco et al. 2016). 
These parameters are in good agreement with 
those ones we obtained from different previous works. 
The only exception is the effective temperature
where the former studies support the lower DR2 value.

Based on our spectra, we give an independent estimate of the atmospheric parameters of V764\,Mon.
We prepared a cumulative median spectrum from the six highest $S/N$ ratio spectra, then we fitted synthetic spectra 
between 5400 and 5900 \text{\AA} to this median spectrum by minimizing $\chi^2$.
The fitting process was similar to what we described for T\,Sex in
detail (Benk\H{o} et al. 2021).
The only difference was that we used Kurucz atmosphere models (Castelli and Kurucz 2003) that better fit our 
parameters instead of the MARCS GES models used for T\,Sex. We applied the synthetic spectral fitting tool 
of iSpec package (Blanco-Cuaresma et al. 2014; Blanco-Cuaresma 2019) for this process.
Different initial value combinations of 
$T_{\mathrm{eff}}$, $\log g$  and [Fe/H] have
chosen from the values in Table 5. The initial micro-turbulent velocity was 
the default value of the algorithm ($v_{\mathrm{mic}}=1.05$~km\,s$^{-1}$). For the macro-turbulent velocity,  the empirical relation offered by the algorithm was accepted. 
During several runs, $v\sin i$ was also a free parameter, with the initial value 188~km\,s$^{-1}$.  
We obtained 199$\pm$20~km\,s$^{-1}$ from the fit. 
Since this value is consistent with our previous result, we fixed $v\sin i$ at 188~km\,s$^{-1}$ for the further runs.
The atmospheric parameters resulting from the fit and their errors are shown in column 6 of Table~5.
As we have seen these values similar to the former estimations but they have rather high standard deviation. 

We completed Table~5 with several calculated 
parameters (italic numbers) using the basic formulae of luminosity, $\log g$ and
the calibrated empirical relation for mass and radius by Torres et al. (2010). For pre-main sequence stars, the latter relation can not be applied.
If we plot our star on the  Hertzsprung-Russell diagram (HRD) using physical parameters from different publications (Fig.~5), 
we find that the estimated position we obtained is the closest to that of EDR3.

\begin{figure}[htb]
\includegraphics[angle=270,width=12cm, trim=1cm 0 1cm 0]{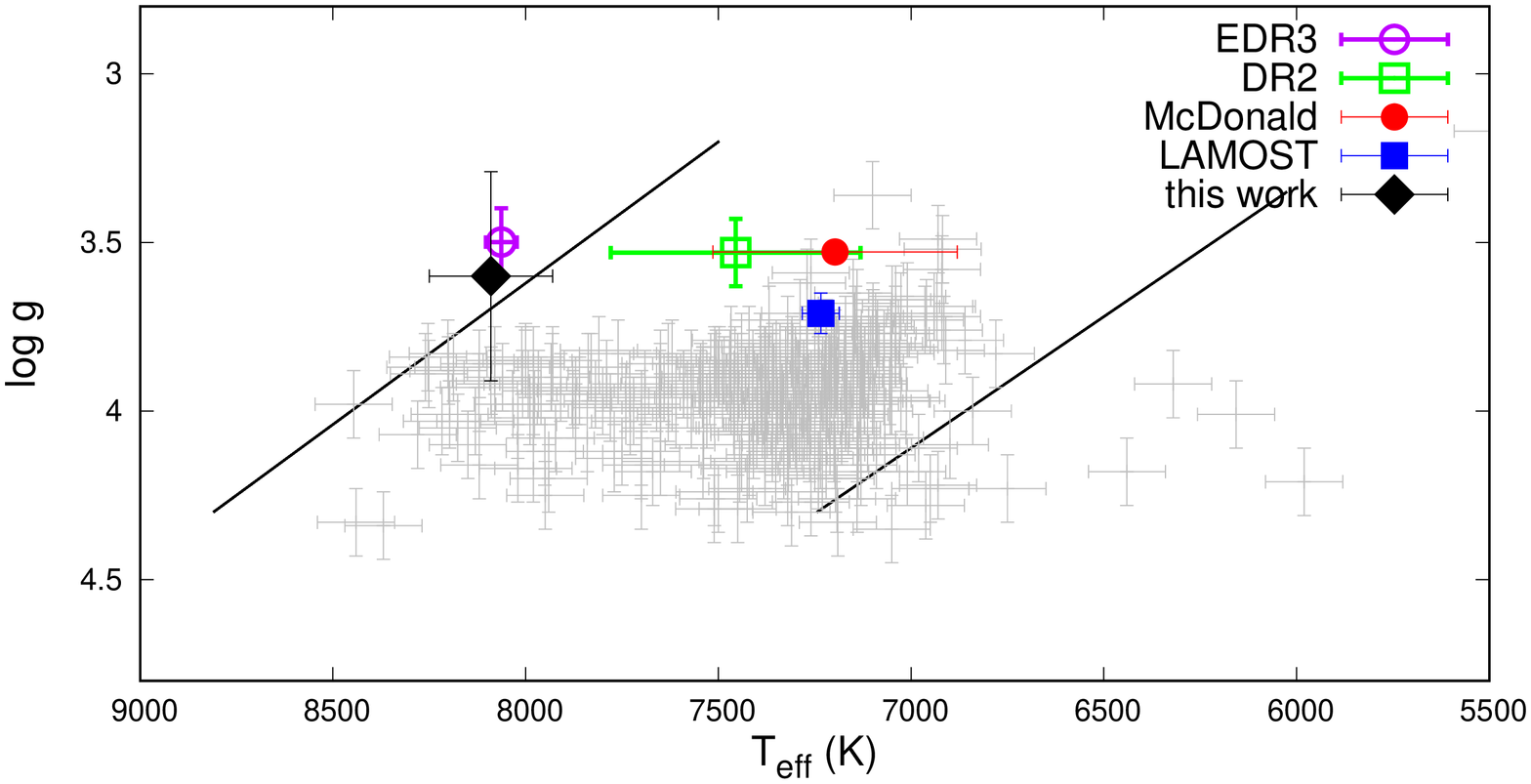}\\
\FigCap{
The position of V764\,Mon in HR diagram from the different estimations listed in Table~5.
Light-grey symbols show normal $\delta$~Scuti stars classified by 
the LAMOST spectroscopic survey (Qian et al. 2018).
The straight line show the instability strip of
classical $\delta$~Scuti stars (Breger and Pamyatnykh 1998)
}
\end{figure}

An additional physical parameter might be estimated.
Searching for regular patterns in the \textit{TESS} pulsation frequencies in Table~2. 
with the SSA method (Papar\'o et al. 2016a,b), we found a frequency sequence of 4 elements consisting of $f_1$, $f_4$, $f_6$ and $f_{12}$. 
The average spacing between the frequencies is 3.24$\pm0.04$~d$^{-1}$.
If we assume this spacing value to be the large separation ($\Delta \nu$), the value of V764\,Mon fits well to the theoretical 
relation between $\Delta \nu/ \Delta \nu_{\odot}$ and $\overline{\rho} / \overline{\rho}_{\odot}$ 
of Su\'arez et al. (2014). (The solar values of the large separation and average density was the same as in Su\'arez et al. 2014 work.) 
The average density of V764\,Mon determined this way ($\overline{\rho} = 0.159$~g / cm$^{-3}$) is typical for a $\delta$~Scuti star.

\subsection{Classification}

To sum up, V764\,Mon is a binary system with a fast rotating main component showing small-amplitude, multiperiodic light variations originating from non-radial pulsation. 
The star is above the main sequence in the HRD and, depending on the parameter estimations,
within or at the edge of the $\delta$~Scuti instability strip (Fig.~5).

This picture is disturbed only by the fact that the dominant period of the star is unusually long compared to 
typical $\delta$~Scuti stars.
The \textit{Kepler} $\delta$~Scuti sample of Balona and Dziembowski (2011) ends at less than 0.2~d.
The most up-to-date Galactic $\delta$~Scuti catalogue (Chang et al. 2013) 
gives a period of $\sim$0.25 days for the edge of the $\delta$ Scuti group. Even though this catalogue contains six stars 
with $P>0.28$~d,
by examining these stars individually, it turns out that they either actually have a shorter 
dominant period, or are not $\delta$~Scuti stars at all. 
However, the LAMOST spectroscopic survey (Qian et al. 2018) identifies a number of $\delta$~Scuti stars up to 
a period of $\sim$0.35 days.

The evolutionary state of V764\,Mon is unclear. Its position above the HRD shows that it is a pre- or
post-main sequence star. $\delta$ Scuti variation of such stars are known 
(see Breger 2007; Zwintz et al. 2014 and references therein).
A distinction between these evolutionary stages based purely on the position in the HRD is not possible because pre- and post-main sequence evolutionary tracks intersect (Breger and Pamyatnykh 1998). Evolutionary identification with detailed asteroseismology would only be possible with accurate mode identification.

\section{HY\,Com}

\subsection{Photometry}

The variability of HY\,Com was discovered by Oja (1981). He determined the period ($P=0.448614$~d) from $UBV$ photoelectic observations, and classified the star as an RRc type pulsating variable. Oja detected a phase variation of the light curve and 
drew the attention to the unusually long pulsation period compared to other RRc variables. He re-observed the star in the 1990s (Oja 1995) and 
found that HY\,Com shows evident short-term period variation of unknown nature. 
Using new photometric data that has become available in the meantime (NSVS, Wo\'zniak et al. 2004; ASAS-3 data between 2002 and 2006, Pojma\'nski et al. 2005), Wils (2008) confirmed the considerable phase/period changes in the star. 
This work showed that 
the photoelectric $V$ light curve, collected between 1994 and 2006, folded with the pulsation period shows only a small scatter, assuming three ''sudden'' period changes in 2001, 2004 and 2005. The shape of the light curve seemed to be stable within the detection error.

Since then, the ASAS-3 photometric data were extended with three more observing seasons (2006/2007, 2007/2008 and 2008/2009). 
For the proper fit of these new observations to
the phase curve of Wils (reproduced in Fig.~6) at least one
additional period change have to be assumed between the last two seasons (2007/2008 and 2008/2009).
The determined periods of the season 2008/2009 is
$P_1=0.448698\pm0.000017$~d, the assumed period change is +11.4~s.

\begin{figure}[htb]
\includegraphics[angle=270,width=11cm, trim=2cm 0 1cm 0]{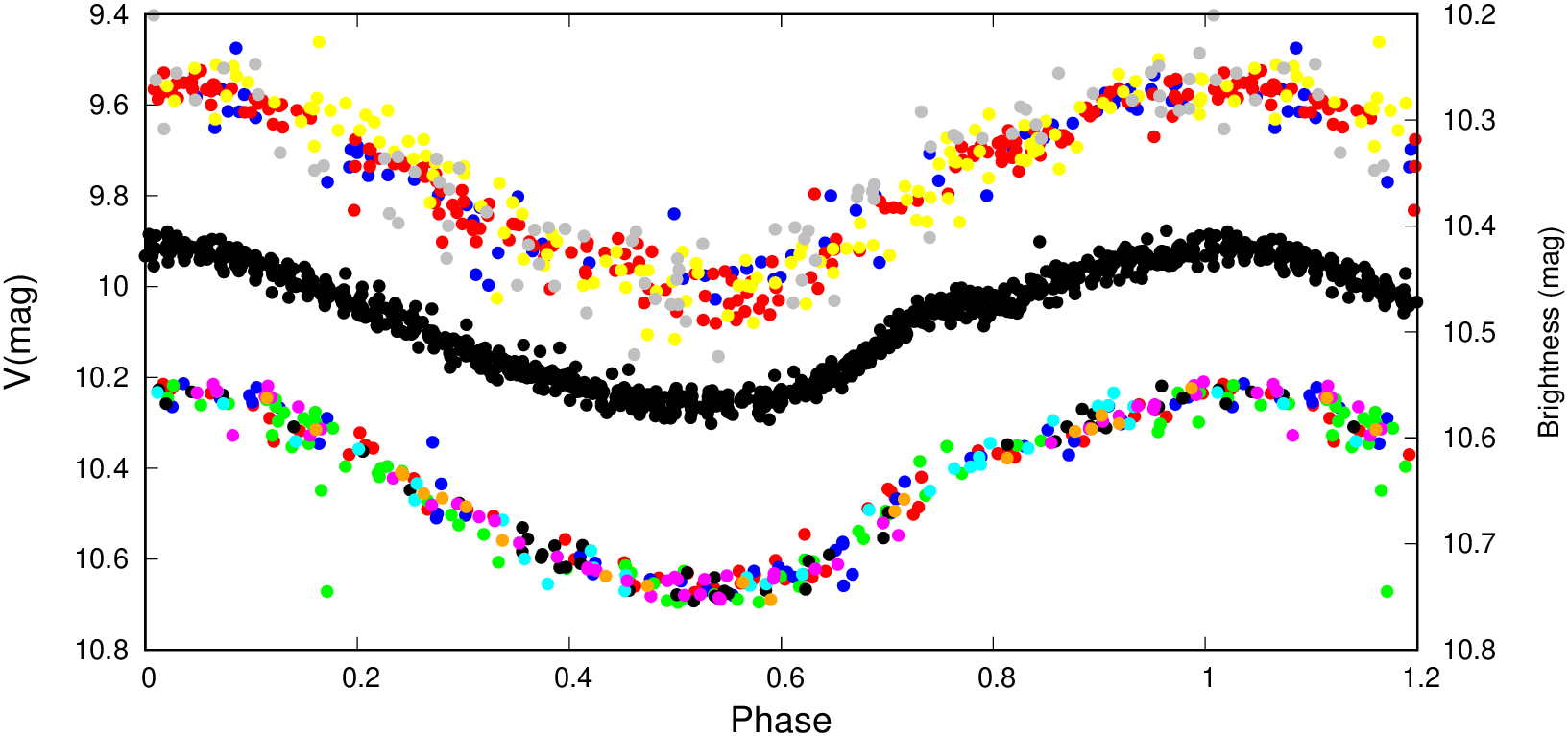}\\
\FigCap{
Phase curves of HY\,Com from the ASAS-3 observations 
(bottom curve), from the Qatar Exoplanet Survey data (middle curve)
and from ASAS-SN survey (top curve). 
The $V$ magnitudes for ASAS and $V-0.7$~mag for ASAS-SN data are indicated in the left axis,
the QES unfiltered magnitudes are shown in the right vertical axis.
Points of ASAS and ASAS-SN data are marked in a different colour 
for each observing season. 
}
\end{figure}

The Qatar Exoplanet Survey (QES, Alsubai et al. 2013, Bramich et al. 2014) also provide us a large amount of photometric data (see Table~1). The instrument of QES consists of five cameras with partially overlapping fields of view. 
HY\,Com was observed by two cameras. The two data sets come from the two cameras differ systematically, mainly around the maximum. Therefore, we used only the data set with smaller standard deviation in our analysis. Even so, we had 775 data points from a 50-day-long interval in 2012.

Analysing the data with the {\sc Period04} Fourier fitting tool, we found two significant frequencies: the main frequency $f_1$ and its third harmonic $3f_1$. This gives a pulsation period of $P_1=0.448783\pm0.000014$~d for 2012, $7$~sec longer than that of the last ASAS-3 season (in 2008/2009).

The ASAS-SN survey (Shappee et al. 2014, Jayasinghe et al. 2019) observed HY\,Com from its Hawaii station at Haleakala Observatory 
in standard $V$-band.
Some 650 data points were observed over five years between 2013 and 2018 (see Table 1). Although the target was saturated, the pipeline handled the issue relatively well (Kochanek et al. 2017).

We analysed the entire data set and the data sets of each season separately. The main frequency $f_1$ was the only significant frequency in all cases.
Based on the periods of individual seasons, we can detect two further significant period changes. The period between seasons 2013/14 and 2015/16 is $P_1=0.448540\pm0.000002$~d which is 21~s shorter than the period of the QES data, while the last two seasons (between 2016/17 and 2017/18) it is again longer with 9.8~s ($P_1=0.448632\pm0.000004$~d) than the previous seasons.

Strong and continuous irregular period changes of RRc stars are rather common. 
This behaviour is especially frequent amongst longer-period RRc stars (see e.g. Jurcsik et al. 2001, 2015).
Details of these period changes (e.g. how abrupt are these) were unclear for a long time.
Continuous \textit{Kepler} measurements have revealed that period changes occur in a few hundred days with 
continuous change (Moskalik et al. 2015, S\'odor et al. 2017), 
and not e.g. with a sudden jump.  However, the physical origins of the phenomenon are still unclear.

\subsection{Spectroscopy}

The spectroscopic observations and reduction were made for HY\,Com with 
the same equipment and 
the same way as we described for V764\,Mon in Sect.~2.3. 
The log of observations is given in Table~3. 
As can be seen from Table~3, the signal-to-noise ratio of the 
HY\,Com measurements is significantly lower than that of the V764\,Mon measurements. The most obvious use of spectroscopic 
data of this quality is to determine the radial velocity curve. This is especially important, since no such data has been published before.

The radial velocities were calculated with 
the same cross-correlating technique as it was done for V764\,Mon.
Gaussian fitting functions were used for determining the radial velocities from the the cross-correlation functions. 
The estimated standard errors of the individual radial velocity values
are $\sim$1\,--\,2~km\,s$^{-1}$.
The radial velocities and their errors are given in Table~6.
\MakeTable{cccccc}{7cm}{Radial velocities of HY\,Com}
{\hline
BJD &  $v_{\mathrm{rad}}$  & $\sigma(v_{\mathrm{rad}})$ & BJD &  $v_{\mathrm{rad}}$  & $\sigma(v_{\mathrm{rad}})$ \\
    (d)  & km\,s$^{-1}$ &  km\,s$^{-1}$        &    (d)  & km\,s$^{-1}$ &  km\,s$^{-1}$        \\
\hline
2457088.50709  & $-49.06$    &   2.31 & 2457090.51150  & $-26.85$  &   1.58\\
2457088.52806  & $-46.04$   &   1.44 & 2457090.53246  & $-27.14$   &   2.12\\
2457088.54903  & $-42.83$   &   1.28 & 2457090.55341  & $-27.53$   &   1.89\\
2457088.57320  & $-41.25$   &   1.37 & 2457090.57530  & $-29.55$  &   2.01\\
2457088.61515  & $-35.69$   &   1.30 & 2457090.59625  & $-32.05$  &   1.74\\
2457088.63712  & $-32.99$  &   1.00 & 2457090.61720  & $-41.75$   &   2.68\\
2457088.65807  & $-33.50$   &   0.77 & 2457091.55324  & $-51.62$  &   1.19\\
2457090.42245  & $-36.12$   &   1.57 &2457091.57640  & $-55.81$  &   1.82\\
2457090.44865  & $-31.27$  &   2.72 & 2457091.59736  & $-58.43$  &   2.46\\
2457090.46960  & $-31.11$  &   1.89 & 2457091.61831  & $-52.83$   &   1.74\\
2457090.49055  & $-28.74$   &   1.77 & & & \\
\hline
\multicolumn{6}{p{7cm}}{
}
}

The radial velocity curve is constructed by folding the radial velocities with the period of  photometry closest to the 
measurements in time (ASAS-SN data, Fig.~7). 
The total amplitude of curve is $\sim$32~km\,s$^{1}$.
Its shape is asymmetric: a proper fit requires four harmonic components of the Fourier solution.
The mean velocity of HY\,Com from this four-element Fourier fit is $-39.9\pm0.7$~km\,s$^{-1}$.
Where different nights are overlapped in phase (around $\phi=0.3-0.5$) a slight
radial velocity shift can be seen which might be caused by some period change. 
\begin{figure}[htb]
\includegraphics[angle=270,width=11cm, trim=1cm 0 1cm 0]{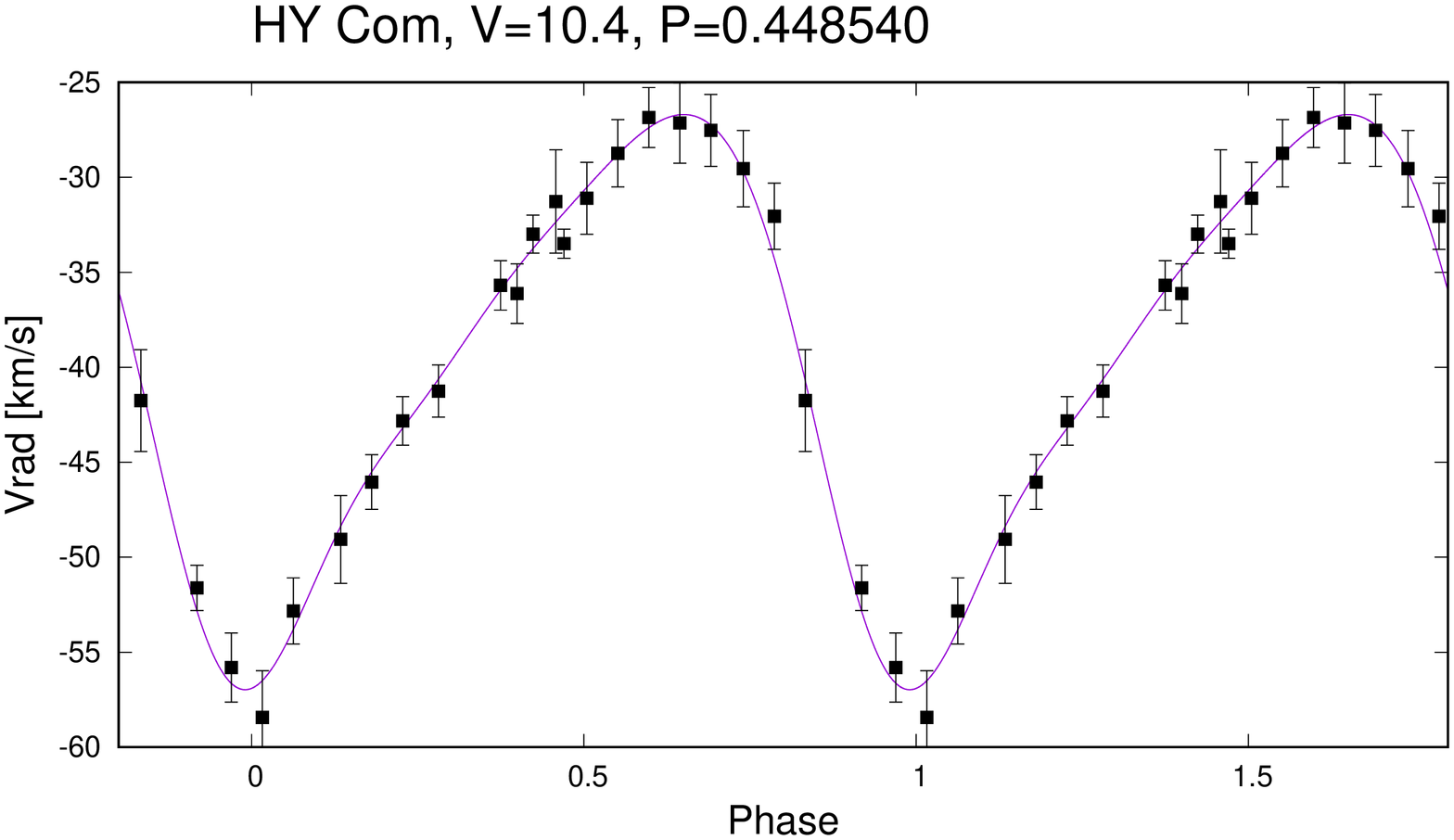}
\FigCap{
Radial velocity curves of HY\,Com.
The error bars show the calculated formal error (typically 1-2~km\,s$^{-1}$). The continuous line shows a four-element Fourier fit.
The initial phase 
($\phi=0=1$) was set to the minimum of the radial velocity curve.
}
\end{figure}

\section{Summary}

On the basis of our spectra and the {\it TESS} photometric data
we showed that the brightest RRc candidate V764\,Mon, which was also a candidate binary RR\,Lyrae star, is, in fact, not an RR\,Lyrae star but a fast rotating $\delta$\,Scuti variable.
Analysing the \textit{TESS} data, we identified ten independent frequencies
that describe the multi-periodic light curve variation.
Our spectra show characteristic signs (time-dependent line-profile distortions) of
non-radial pulsation, which might explain most of the found frequencies in the photometry.
Using spectroscopic cross-correlation functions, we determined the projected rotation velocity 
($v\sin i=188\pm 5$~km\,s$^{-1}$), which shows that V764\,Mon is a fast rotating star. 
We also detected the presence of a companion star, that is, we spectroscopically confirmed the binary nature of V764\,Mon, discovered by the Gaia satellite by astrometric means.
Although the dominant photometric period of V764\,Mon is unusually long (0.29~d), all other photometric and spectroscopic behaviour and parameters confirm its classification as a $\delta$~Scuti pulsator.

Investigating the previously unanalyzed photometric data of HY\,Com, we found that, in agreement with previous studies, 
the shape of the light curve is surprisingly stable, but its phase is not constant over time. 
We need to assume at least four phase changes 
in order to transform the new data into 
a common ephemeris with the older ones.
We also presented here the first complete radial velocity curve of HY\,Com.

\Acknow{
This work was supported by the Hungarian National Research,
Development and Innovation Office by the Grant NN-129075
and the Lend\"ulet Program of the Hungarian Academy of Sciences,
project No. LP2018-7/2018. JMB thanks to Dr M. Papar\'o for her valuable comments.
}

\end{document}